\documentstyle[11pt]{article}
\begin{document}
\begin{titlepage}

\hspace{11cm}IHEP-TH-98-16

\vspace{1cm}

\centerline{\Large \bf Nonperturbative Corrections to One Gluon Exchange }
\centerline{\Large \bf  Quark Potentials 
\footnote{E-mail address: yangjj@bepc3.ihep.ac.cn
}
}

\vspace{1cm}
\centerline{ J.J. Yang$^{a,b,c}$,
H.Q. Shen$^{c}$, G.L. Li$^{a,b}$, T. Huang$^{a,b}$, P.N. Shen$^{a,b,d}$}
\vspace{1cm}
{\small
{
\flushleft{\bf  $~~~$a. China Center of Advanced Science and Technology
 (World Laboratory),}
\flushleft{\bf  $~~~~~~$P.O.Box 8730, Beijing 100080, China}
\vspace{8pt}
\flushleft{\bf  $~~~$b. Institute of High Energy Physics, Academia Sinica,
 P.O.Box 918(4),}
\flushleft{\bf  $~~~~~~$Beijing 100039, China}
\vspace{8pt}
\flushleft{\bf  $~~~$c. Department of Physics, Nanjing Normal  University, 
Nanjing 210097, China}
\vspace{8pt}
\flushleft{\bf  $~~~$d. Department of Physics and Mathematical Physics, the
 University of Adelaide,}
\flushleft{\bf  $~~~~~~$GPO Box 498, Adelaide, South Australia 5001}
}}

\vspace{2cm}

\centerline{\bf Abstract}

\noindent

The leading nonperturbative QCD corrections to the   
one gluon exchange  quark-quark, quark-antiquark and 
$q \bar{q}$ pair-excitation potentials are derived by using 
a covariant form of nonlocal two-quark and two-gluon 
vacuum expectation values. Our numerical calculation indicates that 
the correction of quark and gluon condensates to the quark-antiquark 
potential improves  the heavy quarkonium spectra to some degree.

\end{titlepage}

\baselineskip 18pt

\noindent {\bf 1. Introduction}

\vspace{0.5cm}

A remarkable success of the potential model in explaining hadronic 
spectra and hadronic  properties has been made \cite{Spect,Muller,Gupta}.  
Inspired by QCD,
various potential forms \cite{Potential}, in which both the  
Coulomb potential arising from the one gluon exchange (OGE) mechanism and 
confinement potential are employed,  
can give results which are 
reasonably consistent with the data at the charm and bottom energy
scales within a certain error. 

Along a parallel direction to the QCD sum rule approach \cite{SVZ}, Shen 
et al. \cite{SPN1} derived the leading nonperturbative QCD correction 
to the OGE quark-quark potential and investigated possible effects of the 
nonperturbative contribution in heavy quarkonium spectra. 
In this paper, we will extend 
the nonperturbative calculation of Ref. \cite{SPN1} to 
the nonperturbative correction to the quark-antiquark 
and $q \bar{q}$ pair-excitation potentials 
in a covariant form of the nonlocal two-gluon vacuum expectation value (VEV).

According to some new features which appear in the newly obtained
quark-quark potential including the nonperturbative correction, there
are some intended applications of the modified OGE potentials.
Firstly, the modified quark-quark potential can be used to  study the  
nonperturbative effect in the spectra of $J/\Psi$ and ${\Upsilon}$, 
especially to improve the spin splitting for
these system. Because the ($q\bar{q}$) excitations give significant 
contributions to the nucleon and delta masses, we also hope that modified 
$q\bar{q}$ excitation potential can be employed to 
investigate possible nonperturbative effects in these  
hadronic properties. In addition, 
the nonperturbative effect in decay constants of 
the pseudoscalar mesons can also be considered 
by involving the vacuum condensate effect.
Even some long-standing problems in light  
baryon spectroscopy such as the energy level order between the positive- and 
negative- parity partner states,
in particular, Roper resonance puzzle, and the baryon 
spin-orbit structure puzzle--the absence of spin-orbit effects 
in baryon spectrum can partly be understood along this direction.

In the nonrelativistic reduction of the perturbative OGE potential, 
if the CVC theorem is respected, i.e.
$q^{\mu} \bar{u}_{q}(p_{1}) \gamma_{\mu} u_{q}(p_{1}')=0$ 
with $q=p_1-p^\prime_1$,
the term $q_{\mu}q_{\nu}/q^{2}$ in the  S-matrix for a quark-quark scattering 
vanishes, and the quark-quark interaction is not affected by the 
gauge dependence of the  gluon propagator.

However, the gauge issues  exist when both the 
gluon propagators and the external QCD vacuum gluon fields 
appear in the S-matrix for the nonperturbative quark-quark scattering.
For the gluon propagators, we will 
adopt Landau gauge as done in Refs. \cite{SPN1} and \cite{Larsson} since  
it has been  pointed out in Ref. \cite{Corn} that a Landau gauge calculation 
for an on-shell  amplitude  can yield a correct gauge-invariant result;
For the external QCD vacuum gluon fields, the 
fixed-point gauge \cite {Schwinger} is generally adopted, i.e.,   

\begin{equation}
x_\mu B^{\mu}_a(x)=0.
\end{equation}
Hence the nonlocal two-gluon  VEV is \cite{SVZ,Elias}

\begin{eqnarray}
\langle 0|B^a_\mu(x)B^b_\nu(y)|0 \rangle &=& \frac{1}{4}x^\rho y^\sigma
\langle 0|G^a_{\rho \mu}G^b_{\sigma \nu} |0 \rangle
  + \cdots \cdots  \nonumber \\
  &=& \frac{\delta_{ab}}{48(N_c^2-1)} x^\rho y ^\sigma (g_{\rho \sigma}
  g_{\mu \nu}-g_{\rho \nu} g_{\sigma \mu})
  \langle 0 | G^2 | 0 \rangle +\cdots, \label{BB1}
\end{eqnarray}
where

\begin{equation}
  \langle 0 |G^2 | 0 \rangle=\langle 0|G^a_{\rho \mu}G_a^{\rho \mu} |0 \rangle.
\end{equation}
The expansion of (\ref{BB1}) obviously violates the translational invariance since the 
right hand side(RHS) of (\ref{BB1}) is a function of $xy$ instead of $(x-y)$.

In Ref. \cite{YSL}, we pointed out  that  
there is  an uncertainty  in the 
evaluation of the gluon condensate contribution to a three-point 
nonperturbative vertex provided that the fixed-point gauge of the QCD vacuum  
gluon fields is adopted. We found in Ref. \cite{YCLS} 
that, by using a covariant form of the nonlocal two-gluon VEV, 
the  gluon condensate correction to a 
fermionic three-point vertex can be unambiguously deduced. 

In order to avoid  the ambiguity in the fixed-point gauge, it is 
better to derive the condensate correction to the OGE 
potentials  by means of  
the covariant form of $\langle 0|B^a_\mu(x)B^b_\nu(y)|0 \rangle$ 
given by Bagan et al. \cite{Bagan},
\begin{eqnarray}
\langle 0 | B^a_\mu (x) B^b_\nu (y) | 0 \rangle &=&
- \frac{\delta_{ab}}{(N_c^2-1)}S \left [ (x-y)^2 g_{\mu \nu} 
- \frac{2}{5} (x-y)_\mu 
(x-y)_\nu \right ] + \cdots \cdots, \label{BB9}
\end{eqnarray}
where
\begin{equation}
S=\frac{5 \langle 0|G^2 |0 \rangle}{288}.   \label {S} 
\end{equation}
Deriving the condensate corrections to  the perturbative OGE potentials
by means of the covariant form of nonlocal two-quark and two-gluon VEVs 
is the main purpose of this paper. 

In the following three sections, using the covariant form of nonlocal  
two-gluon VEV \cite{Bagan}, we derive the leading  nonperturbative 
QCD corrections to the perturbative  OGE quark-quark, 
quark-antiquark and  $q \bar{q}$ pair-excitation potentials, respectively. 
Finally, we present a brief discussion and summary with a 
preliminary  numerical result.

\vspace{0.5cm}

\noindent {\bf 2. Quark-Quark Interaction with Correction from 
Nonperturbative QCD}

\vspace{0.5cm}

The  nonrelativistic  reduction method is used to extract
the nonperturbative correction to the OGE potentials.
The Feynman diagrams that will be  used to evaluate the nonperturbative 
correction to  quark-quark interaction are shown in Fig.~1.  
For the scattering of the two quarks of different flavors, 
by using the covariant form of two-gluon VEV of Ref. \cite{Bagan},
the contribution of the Feynman diagram 1(a) to  the
S-matrix can be written as

\begin{eqnarray}
& &S_{1(a)}(p_1,p_2;p_1^\prime,p_2^\prime)=
ig^4[\bar{\psi}^{-}(p_2^\prime) \gamma^\mu \frac{\lambda^a}{2} \psi^{+}(p_2)]
[\bar{\psi}^{-}(p_1^\prime)  \gamma^\nu \frac{\lambda^b}{2} \psi^{+}(p_1)]D^{aa^\prime}_{\mu \mu^\prime}(q) \nonumber \\
& &\times \int d^4k\left 
[\frac{S}{N_c^2-1} \left (g_{\rho \sigma} g_{l m}-\frac{2}{5}
g_{\rho l}g_{\sigma m}\right ) 
\frac{\partial^2}{\partial k_l \partial k_m}\delta^4 (k)\right ]
 \nonumber \\
& &\times \delta_{d d^\prime} f_{a^\prime c d} \left [(2q+k)^\rho g^{\mu^\prime \lambda}
+(-q-2k)^{\mu^\prime} g^{\rho \lambda}
+(k-q)^\lambda g^{\rho \mu^\prime}\right ] 
D^{c c^\prime}_{\lambda \lambda^\prime}(q+k) \nonumber \\
& &\times  f_{c^\prime b^\prime d^\prime} 
\left [(2q+k)^\sigma g^{\lambda^\prime \nu^\prime} 
+(-q+k)^{\lambda^\prime}g^{\nu^\prime \sigma }+ 
(-2k-q)^{\nu^\prime} g^{\lambda^\prime \sigma}\right ] 
 D^{b^\prime b}_{\nu^\prime \nu}(q). \label{S1a}
\end{eqnarray}
Eq.~(\ref{S1a}) can be rewritten as a relation between 
the corresponding quark potential:

\begin{eqnarray}
V_{1(\rm{a})}(q)=\frac{768 \pi \alpha_s N_c  S }{5(N_c^2-1)q^4}
V^{\rm{OGEP}}_{qq}(q),\label{V1a}
\end{eqnarray}
 where $V^{\rm{OGEP}}_{{qq}}(q)$ is the 
usual perturbative quark-quark 
potential arising from the OGE 
mechanism. 
Performing Fourier transformation to $V_{\rm{1(a)}}(q)$ of (\ref{V1a}), we  
obtain the  
contribution of the Feynman
diagram 1(a) to the quark-quark potential 
in the coordinate representation,

\begin{eqnarray}
U_{1(\rm{a})}(x) = \delta (t) \frac{\lambda_1^a \lambda_2^a}{4} \pi \alpha_s^2
\left [A_{3} |\vec{x}|^{3}+A_{1} |\vec{x}|\right ],  \label{V1ax} 
\end{eqnarray}
where

\begin{eqnarray}
A_3=\frac{32N_c S}{5(N_c^2-1)}\left (1+\frac{|\vec{p}|^2}{m_1 m_2}\right ),
\end{eqnarray}

\begin{eqnarray}
A_1&=&\frac{48N_cS}{5(N_c^2-1)m_1 m_2} \left \{ \frac{(m_1 +m_2)^2}{m_1 m_2} 
+\frac{1}{2}\left [3(\vec{\sigma}_1 \cdot \vec{\sigma}_2)
-(\vec{\sigma}_1\cdot \vec{n})(\vec{\sigma}_2\cdot \vec{n}) \right ] \right.\nonumber \\
& &+\left. \frac{1}{2}\left [(2+\frac{m_2}{m_1})\vec{\sigma}_1
+(2+\frac{m_1}{m_2}) \vec{\sigma}_2\right ] \cdot (\vec{x}\times \vec{p})\right \} \nonumber \\
\end{eqnarray}
with $\vec{n}={\vec{x}}/{|\vec{x}|}$. 
Fig.~1(b)  makes no contribution, i.e., $U_{1(\rm{b})}(x)=0$, because the vacuum cannot possess any 
quantum number such as color.

By means of the two-quark VEV \cite{Elias}, 
the contribution of the Feynman diagram 1(c) to the S-matrix can 
be expressed as
\begin{eqnarray}
& &S_{1(\rm{c})}(p_1,p_2;p_1^\prime,p_2^\prime)=-ig^4
[\bar{\psi}^{-}(p_2^\prime)  \gamma^\nu \frac{\lambda^b_2}{2} \psi^{+}(p_2)]
[\bar{\psi}^{-}(p_1^\prime)  \gamma^\mu \frac{\lambda^a_1}{2} \psi^{+}(p_1)] \nonumber \\
& &\times \int d^4k\delta^4 (k) 
 \langle 0|\bar{q}_fq_f |0 \rangle \left [\frac{1}{4N_c}+\frac{m_f}{16N_c} 
 \gamma^\tau \frac{\partial}{\partial k^\tau} \right ]  \gamma^\rho 
 \frac{\lambda_f^{a^\prime}}{2} S(q+k)\nonumber \\ 
 & &\times  \gamma^\sigma \frac{\lambda_f^{b^\prime}}{2} 
 D^{aa^\prime}_{\mu \rho}(q) D^{bb^\prime}_{\nu \sigma}(q),
\end{eqnarray}
where  the  next-to-leading-order term in the full
coefficient of the $\langle \bar{q}q \rangle$ component of the 
nonperturbative two-quark VEV \cite{Elias} is retained.
There exists a similar  expression for Fig.~1(d). It is easy to find that
the contributions of the Feynman diagrams 1(c) and 1(d) to the
effective potential  are the same.  Therefore, we can extract the 
effective potential for the scattering of two quarks of different flavors  
as shown in Fig.~1(c) and Fig.~1(d) in the coordinate representation

\begin{eqnarray}
& &U_{1(\rm{c})}(x)=U_{1(\rm{d})}(x)\nonumber \\ 
&=&\delta (t) \frac{\lambda_1^a \lambda_2^a}{4} \pi \alpha_s^2
\left [C_1 |\vec{x}|+C_{-1}|\vec{x}|^{-1}+C_{-3} |\vec{x}|^{-3}\right. \nonumber \\
& &\left. +\sum\limits_f\left (\tilde{C}_0^{(f)}+\tilde{C}_{-1}^{(f)}|\vec{x}|^{-1}\right ){\rm{e}}^{-m_f |\vec{x}|}  
\right ],
\end{eqnarray}
where

\begin{eqnarray}
C_1=\left (1+\frac{|\vec{p}|^2}{m_1 m_2}\right )\sum  \limits_{f}
\frac{ \langle 0|\bar{q}_fq_f |0 \rangle}{N_c m_f},
\end{eqnarray}

\begin{eqnarray}
C_{-1}&=&\frac{2}{N_c}\sum  \limits_{f}
\frac{ \langle 0|\bar{q}_fq_f |0 \rangle}{m_f}
\left \{ \frac{1}{2m_1 m_2} (\vec{\sigma}_1\cdot \vec{n})(\vec{\sigma}_2\cdot \vec{n})
+\frac{1}{8}\left (\frac{1}{m_1}+\frac{1}{m_2}\right)^2 \right.   \nonumber \\
 & &+\left. \frac{1}{2m_1 m_2}
\left [\left (2+\frac{m_2}{m_1}\right )\vec{\sigma}_1
+\left (2+\frac{m_1}{m_2}\right ) \vec{\sigma}_2\right ] 
\cdot (\vec{x}\times \vec{p})\right \}, 
\end{eqnarray}

\begin{eqnarray}
C_{-3}=\frac{3}{m_1 m_2 N_c} \sum \limits_{f} 
\frac{ \langle 0|\bar{q}_fq_f |0 \rangle}{m_f^3}
(\vec{\sigma}_1\cdot \vec{n})(\vec{\sigma}_2\cdot \vec{n}),
\end{eqnarray}

\begin{eqnarray}
\tilde{C}_{0}^{(f)}&=&-\frac{1}{N_c}
 \langle 0|\bar{q}_fq_f |0 \rangle 
\left \{\frac{1}{m_f^2} \left(1+\frac{|\vec{p}|^2}{m_1 m_2}\right)  
+ \frac{1}{8}\left (\frac{1}{m_1}+\frac{1}{m_2}\right )^2 \right.\nonumber \\
& &+\left. \frac{1}{4m_1m_2}
 \left [(\vec{\sigma}_1 \cdot \vec{\sigma}_2)
-(\vec{\sigma}_1\cdot \vec{n})(\vec{\sigma}_2\cdot \vec{n}) \right ]\right \},
\end{eqnarray}
and

\begin{eqnarray}
\tilde{C}_{-1}^{(f)}&=&\frac{1}{N_c}
\frac{\langle 0|\bar{q}_fq_f |0 \rangle}{m_f}
\left \{ \frac{1}{4m_1 m_2} \left [(\vec{\sigma}_1 \cdot \vec{\sigma}_2)
+ (\vec{\sigma}_1\cdot \vec{n})(\vec{\sigma}_2\cdot \vec{n})\right ]
+\frac{1}{4}\left (\frac{1}{m_1}+\frac{1}{m_2}\right )^2  \right.  \nonumber \\ 
& &-\left.  \frac{1}{2m_1 m_2} \left [\left (2+\frac{m_2}{m_1}\right )\vec{\sigma}_1
+\left (2+\frac{m_1}{m_2}\right ) \vec{\sigma}_2\right ] 
\cdot (\vec{x}\times \vec{p})\right \}. 
\end{eqnarray}
The total quark-quark effective potential is finally  obtained by 
summing up  the contributions of all the 
corresponding diagrams including 
perturbative and  nonperturbative ones:

\begin{eqnarray}
U_{qq}(x)=U_{qq}^{\rm{OGEP}}(x)+U_{qq}^{\rm{NP}}(x) \label{U1}
\end{eqnarray}
where $U_{qq}^{\rm{NP}}(x)$, the nonperturbative correction 
to the perturbative
quark-quark potential  due to the quark and gluon condensates, can be expressed as 

\begin{eqnarray}
U_{qq}^{\rm{NP}}(x)&=&U_{1(\rm{a})}(x)+ U_{1(\rm{b})}(x)+ U_{1(\rm{c})}(x)+ U_{1(\rm{d})}(x) \nonumber \\
&=& \delta (t) \frac{\lambda_1^a \lambda_2^a}{4} \pi \alpha_s^2 
\left [A_3 |\vec{x}|^3+\left (A_1+ 2C_1\right ) |\vec{x}| 
+2 C_{-1} |\vec{x}|^{-1} \right. \nonumber \\  
& &\left. + 2 C_{-3} |\vec{x}|^{-3}
+2\sum\limits_f\left (\tilde{C}_0^{(f)}
+\tilde{C}_{-1}^{(f)}|\vec{x}|^{-1}\right ){\rm{e}}^{-m_f |\vec{x}|}\right ].\label{U1NP} 
\end{eqnarray}
Formally, Eq.~(\ref{U1}) holds not only for the $qq$-, but also 
for the $q\bar{q}$- and $\bar{q}\bar{q}$-interactions. Note, however,
that the color generators for an antiquark are given by $-{\lambda}^T$, i.e.,

\begin{eqnarray}
U^{\rm{Direct}}_{q\bar{q}}(x)&=& U_{qq}(x)
|_{\lambda_1^a \lambda_2^a\rightarrow -\lambda_1^a (\lambda_2^a)^{T}}
\end{eqnarray}
and 
\begin{eqnarray}
U_{\bar{q}\bar{q}}(x)&=& U_{qq}(x)
|_{\lambda_1^a \lambda_2^a \rightarrow (\lambda_1^a)^{T} 
(\lambda_2^a)^{T}}.
\end{eqnarray}

For an interaction between a quark and an antiquark, if 
the quark and antiquark are of the same kind of quark fields,
then not only the direct scattering but also the 
annihilation mechanism should  be taken 
into account. The detail  discussion is given  in the following  section.

\vspace{0.5cm}

\noindent {\bf 3. Quark-Antiquark Annihilation 
with Correction from Nonperturbative QCD}

\vspace{0.5cm}

In the above section, the direct interaction potential between 
the quark and antiquark was given. When the quark and antiquark 
have  the same kind of flavor, 
then the annihilation of a quark and an antiquark is  possible. The 
annihilation diagrams including the 
nonperturbative correction shown in Fig.~2  should  be further considered.
We can obtain the contributions of these Feynman diagrams to the 
quark-antiquark annihilation potential  by means of 
the same procedure as above. 
Taking Fig.~2(a) as an example, the S-matrix for this diagram  
can be obtained by making 
the substitutions
\begin{eqnarray}
\bar{\psi}^{-}(p_2^\prime) \rightarrow \bar{\psi}^{+}(p_2), 
{\psi}^{+}(p_2) \rightarrow {\psi}^{-}(p_2^\prime), 
m_1=m_2=m
\end{eqnarray} 
in Eq.~(\ref{S1a}).
In the calculation of color part and spin part in the S-matrix for 
diagrams in Fig.~2, 
it is valuable to note that 
\begin{eqnarray}
\sum\limits_{a=1}^{N^2-1} (\lambda^a)_{\beta \alpha} (\lambda^a)_{\alpha^\prime \beta^\prime}
=-\frac{1}{N}
\sum\limits_{a=1}^{N^2-1} (\lambda^a)_{\alpha^\prime \alpha} (\lambda^a)_{\beta \beta^\prime} 
+\frac{2(N^2-1)}{N^2} \delta_{\alpha^\prime \alpha}\delta_{\beta \beta^\prime} 
\end{eqnarray}  
for {\bf{SU}}($N$) generators $\lambda^a(a=1,2,\cdots, N^2-1)$, with $N$=3, 2 
for the  color and spin generators, respectively.
Thus, the effective potential 
for Fig.~2(a) can be obtained directly as follows,

\begin{eqnarray}
V_{\rm{2(a)}}(q)&=& 
\frac{768 \pi \alpha_s N_c S }{5(N_c^2-1)(p_1+p_2)^4}V^{\rm{Ann}}_{q\bar{q}}(q),\label{V2a}
\end{eqnarray}
where
\begin{eqnarray}
V^{\rm{Ann}}_{q\bar{q}}(q)&=&
\frac{4\pi \alpha_s }{(p_1+p_2)^2} 
\left [\frac {(\lambda_1-\lambda_2^T)^2}{8N_c}\right ]
\left [ \frac{(1- \vec{\tau}_1 \cdot \vec{\tau}_2)}{2}\right ] 
 \left \{ \frac{(\vec{\sigma}_1+ \vec{\sigma}_2)^2}{4}\right. \nonumber \\ 
& & \times \left [1-\frac{1}{6m^2}(\vec{q}~^2
+{\vec{q}~^\prime}^2)\right ] 
 -\frac{1}{2m^2}\left [(\vec{\sigma}_1 \cdot \vec{q})
(\vec{\sigma}_2 \cdot \vec{q})
+(\vec{\sigma}_1 \cdot \vec{q}~^\prime)
(\vec{\sigma}_2 \cdot \vec{q}~^\prime) \right. \nonumber \\
& &\left. \left. -\frac{1}{3} \vec{\sigma}_1 \cdot \vec{\sigma}_2
(\vec{q}~^2+{\vec{q}~^\prime}^2)\right ] \right \}. \label{VAnnq}
\end{eqnarray}
where, $\vec{q}$ and $\vec{q}^{~\prime}$ are relative momenta between 
quarks and antiquarks
in the initial and final states, respectively. The isospin factor 
$(1- \vec{\tau}_1 \cdot \vec{\tau}_2)/2$ in
(\ref{VAnnq}) is introduced by considering the fact that the potential has 
a nonvanishing value only for isospin $T=0$ state of a quark and antiquark 
pair, which corresponds to the gluon quantum number.
Performing Fourier transformation to $V_{2\rm{(a)}}(q)$ yields
\begin{eqnarray}
U_{2(\rm{a})}(x) &=& 
\frac{48  \pi \alpha_s N_c S }{5(N_c^2-1)m^4} U^{\rm{Ann}}_{q\bar{q}}(x),\label{U2a}
\end{eqnarray}
where  
$ U^{\rm{Ann}}_{q\bar{q}}(x)$,  
the perturbative  $\rm{q\bar{q}}$ pair-annihilation  potential 
in the coordinate representation, is,

\begin{eqnarray}
U^{\rm{Ann}}_{q\bar{q}}(x)&=& \delta (t) \frac{\alpha_s}{4} \frac{\pi}{16N_cm^2}
(\lambda_1-\lambda_2^{\rm{T}})^2(1- \vec{\tau}_1 \cdot \vec{\tau}_2)
 \nonumber \\
& & \times\left \{ (\vec{\sigma}_1+ \vec{\sigma}_2)^2\left (1-\frac{1}{3m^2} 
\vec{\nabla}^2 \right )\delta (\vec{x})
-\frac{4}{m^2}\left [(\vec{\sigma}_1 \cdot \vec{\nabla})(\vec{\sigma}_2 \cdot \vec{\nabla}) \right. \right.\nonumber \\
 & &\left. \left.-\frac{1}{3} \vec{\sigma}_1 \cdot \vec{\sigma}_2 \vec{\nabla}^2 \right ] \delta(\vec{x})\right \}.
\end{eqnarray}
The nonperturbative 
contributions of Fig.~2(b)-(d) may also be obtained by means of the same 
procedure as above.
The  total ${q \bar{q}}$-pair annihilation potential can be obtained by 
summing up the contributions of all diagrams including nonperturbative 
ones in  Fig.~2 and the corresponding perturbative one, 
\begin{eqnarray}
U^{\rm{Ann}(\rm{Total})}_{q\bar{q}}(x)=U^{\rm{Ann}}_{q\bar{q}}(x)
+ U^{\rm{Ann(NP)}}_{q\bar{q}}(x)
\end{eqnarray}
where
\begin{eqnarray}
U^{\rm{Ann(NP)}}_{q\bar{q}}(x) &=& \frac{\pi \alpha_s}{m^2}
\left \{ \frac{48  N_c S}{5(N_c^2-1)m^2}
+ \frac{1}{N_c}
\sum  \limits_{f}
\frac{m_f (8m^2-m_f^2) \langle 0|\bar{q}_fq_f |0 \rangle}{(4m^2-m_f^2)^2}
 \right \}
 U^{\rm{Ann}}_{q\bar{q}}(x).
\end{eqnarray}
Therefore, the  effective potential between a quark and an antiquark of the same flavor may
be expressed as the summation of the direct and annihilation potentials:
\begin{eqnarray}
U_{q\bar{q}~~\rm{ of ~~the ~~same ~~flavor}}(x)=
U^{\rm{Direct}}_{q\bar{q}}(x)
+U^{\rm{Ann(Total)}}_{q\bar{q}}(x) 
\end{eqnarray}

\vspace{0.5cm}

\noindent {\bf 4. Nonperturbative QCD Corrections to the 
$\rm{q\bar{q}}$ Pair-Excitation Potential}

\vspace{0.5cm}

Now, we turn to the nonperturbative QCD  correction to a ${q\bar{q}}$ pair-
excitation potential. By means of the same  procedure as  above, 
we can calculate the contribution of Fig.~3 to the 
effective potential. The result for Fig.~3(a) reads

\begin{equation}
V_{3(\rm{a})}(q)=\frac{768 \pi \alpha_s N_c S }{5(N_c^2-1)q^4}
V^{q\rightarrow q q\bar{q}}(q),\label{V3aq}
\end{equation}
where $V^{q\rightarrow q q\bar{q}}(q)$, the usual perturbative quark 
excitation potential in the OGE 
approximation, is \cite{Henley}  

\begin{eqnarray}
V^{q\rightarrow q q\bar{q}}(q) &=& 
\frac{\lambda^a_1 \lambda_2^a}{4} 4 \pi \alpha_s \frac{1}{q^2} 
\left [ \frac{1}{2} \left (\frac{1}{m_1}+ \frac{1}{m_2}\right ) 
\vec{q} \cdot \vec{\sigma}_2\right.
 \nonumber \\
& &\left.  -\frac{i}{2m_1}
\vec{q}\cdot (\vec{\sigma}_1\times \vec{\sigma}_2)+
\frac{ \vec{p}_1 \cdot \vec{\sigma}_2 }{m_1} \right ]\label{Vqqqq}
\end{eqnarray}
with $\vec{q}=\vec{p}_1^{~\prime}-\vec{p}_1$.
As suggested in Ref.~\cite{Henley}, we adopt two different approximations, i.e.,
for $q^2=\omega_q^2-\vec{q}^{~2}$,
$\omega_q=0$ (case A), and $\omega_q=2m_2$ with $\vec{q}\simeq 0$ (case B).
From the Fourier transformation, the expression for the transition potential
(\ref{Vqqqq}) in the coordinate representation can be written as 
\begin{eqnarray}
U^{\rm{(A)}q\rightarrow q q\bar{q}}(x)&=-&\delta(t) i \alpha_s
\frac{\lambda^a_1 \lambda_2^a}{4} \frac{1}{2 |\vec{x}|} 
\left \{ \left [\left (\frac{1}{m_1}+\frac{1}{m_2}\right )\vec{\sigma}_2 \right.\right. \nonumber \\
 & &-\left.\left. \frac{i (\vec{\sigma}_1 \times \vec{\sigma}_2)}{m_1}\right ]\cdot
\frac{\vec{x}}{|\vec{x}|^2}
-\frac{2i \vec{\sigma}_2 \cdot \vec{p}_1}{m_1} \right \},
\end{eqnarray}
and
\begin{eqnarray}
U^{\rm{(B)}q\rightarrow q q\bar{q}}(x)&=-&\frac{i \delta(t)}{2m_2^2} 
\frac{\lambda^a_1 \lambda_2^a}{4} \pi \alpha_s
\left \{ \nabla _{\vec{x}}\cdot \left [\frac{\vec{\sigma}_2}{m_1}
+\frac{\vec{\sigma}_2}{m_2} \right.\right. \nonumber \\  
 & &- \left. \left. \frac{i (\vec{\sigma}_1 \times \vec{\sigma}_2) }{m_1}\right ] \delta (\vec{x})
+\frac{2i \vec{\sigma}_2 \cdot \vec{p}_1}{m_1} \delta (\vec{x})\right \}, 
\end{eqnarray}
in case A and case B, respectively.
In case A, the expression for (\ref{V3aq}) in the coordinate representation
is 
\begin{eqnarray}
U_{3(\rm{a})}^{\rm{(A)}}(x)=\delta (t) \frac{\lambda^a_1 \lambda_2^a}{4}
4 \pi \alpha_s^2[D_3|\vec{x}|^3 + D_{2} |\vec{x}|^{2}],
\end{eqnarray}
with

\begin{eqnarray}
D_3=-\frac{8 N_cS}{5(N_c^2-1)m_1} \vec{\sigma}_2 \cdot \vec{p}_1,
\end{eqnarray}

\begin{eqnarray}
D_2=\frac{12 N_cS}{5(N_c^2-1)} 
\left [\frac{\vec{n}\cdot (\vec{\sigma}_1\times \vec{\sigma}_2)}{m_1}  
+i(\frac{1}{m_1}+ \frac{1}{m_2}) (\vec{\sigma}_2 \cdot \vec{n})\right ],
\end{eqnarray}
In case B, one can easily obtain
\begin{eqnarray}
U_{3(\rm{a})}^{(B)}(x)=\frac{48 \pi \alpha_s N_c S}{5m_2^4(N_c^2-1)}
U^{\rm{(B)q\rightarrow q q\bar{q}}}(x).
\end{eqnarray}

Similarly, the  potential for Fig.~3(b) does not give any 
contribution, i.e., $U_{3(\rm{b})}^{\rm{(A/B)}}(x)=0$. 
The  potential for Fig.~3(c) and that for Fig.~3(d) are the same, and turn   
out to be 

\begin{eqnarray}
& &U_{3(\rm{c})}^{(A)}(x) = U_{3(\rm{d})}^{(A)}(x)\nonumber \\
&=&\delta (t) \frac{\lambda^a_1 \lambda_2^a}{4}
4 \pi \alpha_s^2\left [ F_{1} |\vec{x}|+ F_{0} +
\sum\limits_f \tilde{F}_{0}^{(f)}  {\rm{e}}^{-m_f |\vec{x}|}\right ],
\end{eqnarray}
in case A,
with
\begin{eqnarray}
F_1=\frac{\vec{p}_1 \cdot \vec{\sigma}_2}{4N_c m_1}
\sum  \limits_{f}
\frac{ \langle 0|\bar{q}_fq_f |0 \rangle}{m_f},
\end{eqnarray}

\begin{eqnarray}
F_{0} &=& \frac{i}{8N_c} 
\sum  \limits_{f}
\frac{ \langle 0|\bar{q}_fq_f |0 \rangle}{m_f} 
\left [ \frac{i}{m_1} \vec{n} \cdot (\vec{\sigma}_1 \times \vec{\sigma}_2)\right.\nonumber \\
& &-\left. \left (\frac{1}{m_1}+\frac{1}{m_2}\right )(\vec{n}\cdot \vec{\sigma}_2)
\right ], 
\end{eqnarray}
and
\begin{eqnarray}
\tilde{F}_0^{(f)}&=&\frac{\vec{p}_1 \cdot \vec{\sigma}_2}{8N_c m_1}
\frac{ \langle 0|\bar{q}_fq_f |0 \rangle}{m_f^2}\nonumber \\
& &- \frac{i}{16N_c} 
\frac{ \langle 0|\bar{q}_fq_f |0 \rangle}{m_f} 
\left [ \frac{i}{m_1} \vec{n} \cdot (\vec{\sigma}_1 \times \vec{\sigma}_2)\right.\nonumber \\
& &-\left.  (\frac{1}{m_1}+\frac{1}{m_2})(\vec{n}\cdot \vec{\sigma}_2)
\right ]. 
\end{eqnarray}
In case B, 
\begin{eqnarray}
& &U_{3(\rm{c})}^{\rm{(B)}}(x)=U_{3(\rm{d})}^{\rm{(B)}}(x) \nonumber \\
&=& \frac{\pi \alpha_s}{2N_c m_2^2}
\sum  \limits_{f}
\frac{m_f \langle 0|\bar{q}_fq_f |0 \rangle}{(4m_2^2-m_f^2)}
\left [1+\frac{m_f^2}{2(4m_2^2-m_f^2)}\right ] U^{(B)q\rightarrow q q\bar{q}}(x).
\end{eqnarray}
Therefore, the total transition potential in case A (or B) is 
\begin{eqnarray}
U^{\rm{(A/B) q\rightarrow q q\bar{q}}}_{\rm{Total}}(x)= 
U^{\rm{(A/B)q\rightarrow q q\bar{q}}}(x)
+U^{\rm{(A/B) q\rightarrow q q\bar{q}(NP)}}(x),
\end{eqnarray}
$U^{\rm{(A/B) q\rightarrow q q\bar{q}(NP)}}(x)$, the 
nonperturbative correction to the perturbative
$\rm{q \bar{q}}$-pair excitation potential from all 
diagrams shown in Fig.~3  is

\begin{eqnarray}
U^{\rm{(A) q\rightarrow q q\bar{q}(NP)}}(x)&=& U_{3(\rm{a})}^{(\rm{A})}
+ U_{3(\rm{b})}^{\rm{(A)}}+ U_{3(\rm{c})}^{(\rm{A})}+ U_{3(\rm{d})}^{(\rm{A})}\nonumber \\
&=& \delta (t) \frac{\lambda^a_1 \lambda_2^a}{4}
4 \pi \alpha_s^2\left [D_3 |\vec{x}|^{3}+D_2 |\vec{x}|^{2}
+ 2F_{1} |\vec{x}| \right. \nonumber \\
& &+\left.   2F_0 + 2 \sum\limits_f \tilde{F}_{0}^{(f)} 
{\rm{e}}^{-m_f |\vec{x}|}\right ], \label{U2ANP}
\end{eqnarray}
in case A, or
\begin{eqnarray}
U^{\rm{(B)q\rightarrow q q\bar{q}(NP)}}(x)&=& U_{3(\rm{a})}^{(B)}
+ U_{3(\rm{b})}^{\rm{(B)}}+ U_{3(\rm{c})}^{\rm{(B)}}
+ U_{3(\rm{d})}^{\rm{(B)}} \hspace{2cm} \nonumber \\
&=&\frac{\pi \alpha_s}{m_2^2}
\left \{ \frac{48 N_c S}{5(N_c^2-1)m_2^2}
+ \frac{1}{N_c}
\sum  \limits_{f}
\frac{m_f \langle 0|\bar{q}_fq_f |0 \rangle}{(4m_2^2-m_f^2)}
\left [1+\frac{m_f^2}{2(4m_2^2-m_f^2)}\right ] \right \} \nonumber \\
& &\times U^{\rm{(B)q\rightarrow q q\bar{q}}}(x),
\end{eqnarray}
in case B.

\vspace{0.5cm}

\noindent{\bf 5. Some Preliminary Numerical Results}

\vspace{0.5cm}

In order to provide new insight into the effect of 
the condensate correction on the quark-antiquark potential, 
we expect the  obtained  potential including the  nonperturbative  effect
can be employed to improve the hadronic  spectra 
and hadronic properties of $J/{\Psi}$ and ${\Upsilon}$ families,
for example the spin splitting between
1$^{3}S_{1}$ and 1$^{1}S_{0}$ could be one of the 
sensitive quantities for the correction. 
Here we give the result for $J/{\Psi}$ family, the nonperturbative  
effect in ${\Upsilon}$ family is not presented since it is 
qualitatively similar to that of $J/{\Psi}$ family.

\vspace{0.5cm}

Similar to Ref. \cite{SPN1}, the Cornell potential 
$U^{C}=-4\alpha_{s}/3r~+~\kappa r$ with corresponding parameters 
$\alpha_{s}$ and $\kappa$ \cite{Lich}, which gave
the best fit to the $J/\Psi$ and $\Upsilon$ family data, is adopted as a
basic condition, and the values of vacuum condensates are taken 
from Ref. \cite{SVZ}. 
The total potential 
together with the phenomenological
linear confinement can be written in the following form:
$$
U(r)=U^{C}+U^{Corr}_{1}(r)+U^{Corr}_{2}(r)
+U^{Corr}_{3}(r),
$$
where $U^{Corr}_{1}$ is the correction from the non-trivial physical
vacuum condensates, while $U^{Corr}_{2}$ and $U^{Corr}_{3}$ 
(see Ref. \cite{SPN1}) are the Breit-Fermi corrections to the 
Coulomb and linear confinement terms, respectively. 

The numerical calculation is performed in the following way. The
Cornell potential is considered as the dominant  part of the
potential, and the corresponding parameters $\alpha_s$ and $\kappa$
are determined before adding in the corrections. The values of $\alpha_s$
and $\kappa$ are 0.381 and 0.182$GeV^{2}$, respectively \cite{Lich}.
And then, the newly derived corrections
due to the non-vanishing vacuum condensates as well
as the Breit-Fermi corrections are treated as a perturbation adding onto 
the dominant part. The resultant values for the $c\bar{c}$ 
system are tabulated in Table 1. 

\vspace{0.5cm}
\newpage

\centerline{Table 1 $c\bar{c}$ system}

\vspace{0.3cm}

\begin{footnotesize}
\begin{center}
\begin{tabular}{|c|c|c|c|c|c|c|}
\hline
 & $~~~~~~$ & $~~~~~~$ & $~~~~~~$ &
   $~~~~~~$ & $U^{C}+U^{Corr}_{2}$  &
   $~~~~U^{C}+U^{Corr}_{2}~~~~$   \\
 & exp't.  & $U^{C}=$ & $U^{C}+U^{Corr}_{2}$ &
  $U^{C}+U^{Corr}_{2}$ & $U^{Corr}_{3}+U^{Corr}_{1}$ &
   $U^{Corr}_{3}+U^{Corr}_{1}$ \\
 &  & ${-4\alpha_s\over 3r}+\kappa r$ & 
  &+$U^{Corr}_{3}$ & $ (U^{Corr}_{1}$ in~ Ref.\cite{SPN1})  
  & $(U^{Corr}_{1}$ of ours)  \\
 &  &  &  & $\beta=0.6$ & $\beta=0.6$  & $\beta=0.6$ \\
\hline
$1^1s_0$ & 2978.8$\pm$1.9 & 3074.0 & 3010.8 & 3046.9 & 2979.3 & 2981.3 \\
\hline
$2^1s_0$ & 3594.0$\pm$5.0 & 3662.1 & 3625.0 & 3646.4 & 3446.3 & 3466.5 \\
\hline
$1^3s_1$ & 3096.88$\pm$0.04 & 3074.0 & 3095.1 & 3083.0 & 3090.6 & 3094.1 \\
\hline
$2^3s_1$ & 3686.00$\pm$0.09 & 3662.1 & 3674.5 & 3667.3 & 3493.1 & 3516.1 \\
\hline
$1^3p_0$ & 3415.1$\pm$1.0 &        & 3440.8 & 3410.5 & 3321.2 & 3330.7 \\
\hline
$1^3p_1$ & 3510.53$\pm$0.12 &($1P_{c}$)3497.1& 3489.3 & 3481.7 & 3395.1 &
3405.4\\ \hline
$1^3p_2$ & 3556.17$\pm$0.13 &        & 3514.4 & 3531.8 & 3445.1 & 3458.2 \\
\hline
$E_{20}$ & 141.07          &  0     & 73.6  & 121.34  & 123.9   & 127.5 \\
\hline
$E_{21}$ & 45.64           &  0     & 25.1 & 50.1   & 50.0   & 52.7 \\
\hline
$\Delta_{ss}^{(1)}$ & 118.08     &  0    & 84.3  & 36.1 & 111.3  & 112.8 \\
\hline
$\Delta_{ss}^{(2)}$ &  92.0     &  0     & 49.5  & 20.9  & 46.8  & 49.6 \\ \hline

\end{tabular}
\end{center}
\end{footnotesize}

\vspace{0.3cm}
{\footnotesize
In this table, $E_{20}\equiv M_{1^3p_2}-M_{1^3p_0}$,
$E_{21}\equiv M_{1^3p_2}-M_{1^3p_1}$, $\Delta_{ss}^{(1)}\equiv M_{1^3s_1}-
M_{1^1s_0}$ and $\Delta_{ss}^{(2)}\equiv M_{2^3s_1}-M_{2^1s_0}$.} 
$\beta$, which characterizes the fraction of the confinement potential 
which comes from vector exchange, is taken as 0.6 \cite{SPN1}.

{\footnotesize
The experimental data are taken from "Partical Physics Booklet", July 1994,
Partical Data Group.}

\vspace{1cm}

\vspace{0.5cm}

\noindent {\bf 6. Discussion and Summary}

\vspace{0.5cm}

To have a deeper understanding of the hadronic structure,  
there have been various ways to modify the potential.  
As shown by Gupta et al. \cite{Gupta}, Fulcher \cite{Fulcher} 
and Pantaleone et al. \cite{Pant}, the nonperturbative condensate  and 
the perturbative closed-loop corrections are two comparable effects,
because these two kinds of modifications are in the same order of 
$\alpha_s$. Hence, in a complete analysis, nonperturbative condensate 
corrections  should be taken into account. 

The uncertainty due to the  violation of the translational invariance 
in the fixed-point gauge  is unavoidable.
So as to  include the nonperturbative corrections in a more 
self-consistent way  in the nonrelativistic potential reduction,
we used the covariant two-gluon VEV to 
evaluate the  nonperturbative correction to the OGE 
potentials.

In the resultant potentials, some new  features  appear:
A linear term,  which can play a
role of the regular confinement potential, results from the 
condensate corrections  of  quarks and  gluons to the  gluon propagator;
the  modification  of the  gluon condensate to  the gluon  
propagator can offer a $|\vec{x}|^3$ term;  
The Yukawa-type term $|\vec{x}|^{-1} {\rm{e}}^{-m_f |\vec{x}|}$ in 
(\ref{U1NP}) comes  from the nonzero 
quark condensate and may somewhat provide 
an interaction at longer range, such as part of the effects brought 
about by pseudoscalar meson exchange; 
the nonzero quark and gluon condensates can lead to the modification to the
spin-orbital and tensor interactions, which may affect the  
baryon spectrum and  scattering phase shift.
It is noteworthy that the nonperturbative correction can  
not be expected to  include as much as a purely phenomenological ansatz, 
because this framework is an extrapolation from 
perturbative region.  However, it does shed light on the physical 
picture and enrich our understanding of hadronic structure and 
the underlying mechanisms which determine how quarks are 
bound into hadrons.

As a preliminary numerical result, the hadronic 
spectra of $J/\Psi$ family 
and the spin splitting for
this system were calculated by using the modified 
quark-antiquark potential, which indicates that the newly  
introduced corrections does lead to some improvements in the 
description of those  hadronic observables to some degree.
In the present work, only  the lowest-dimension quark 
and gluon condensate corrections are taken into account, 
thus our calculation is reliable up to the 
intermediate range and it can only provide a description of the 
interaction between  the quark and antiquark at the distance 
of $r$($r \simeq 1$ fm). 
For a longer distance scale, higher 
order condensates should be taken into account. 

In summary, it is the goal of this paper to extend the OGE 
potentials for the pure perturbative region 
to the intermediate range in QCD and to investigate the gauge dependence of 
the QCD vacuum gluon fields. 
Here  we only give an example to calculate them within  the lowest
condensate approximation and the result seems to be better than 
those obtained in the fixed-point gauge of the QCD vacuum gluon fields.
In order to improve the numerical result and give a reliable 
result for the higher excited states, the higher-dimension 
condensates should be included.
Physically relevant results, such as the effective quark-quark interaction 
potential,  should be gauge-independent. 
The differences between the present result of quark-quark interaction 
and that of Ref. \cite{SPN1} indicate that there are still  
gauge issues to be clarified.
For this reason, we should  fulfill
the construction of description in Lorentz gauge by including the 
ghost propagators and ghost condensates  as well as  
the $\langle B^2 \rangle$ condensates  
according to the Slavnov-Taylor identities (STI) \cite{Taylor}. 
Further studies, such as the intended applications of the 
modified OGE potentials mentioned in the introduction and the 
improvement of description in Lorentz gauge, are 
in progress. Along this direction, we hope that the 
translational invariance and the gauge issues in the nonperturbative 
calculation can be clarified simultaneously.

\begin{center}
{\bf Acknowledgements} 
\end{center}
One of the authors (J.J.Y.) would like to thank 
professor Z. Y. Zhang and Dr. Y. B. Dong for helpful discussions.
This work was supported in part 
by  the  Natural Science Foundation of China grant no 19775051 and 
the Natural Science Foundation of Jiangsu Province, China.

\newpage

\vspace{3cm}

\section*{Figure Captions}

\vspace{0.6cm}

\mbox{}

\noindent
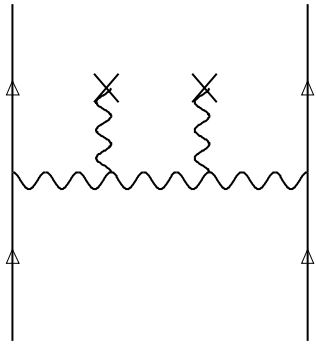
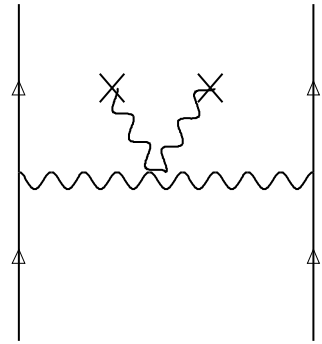
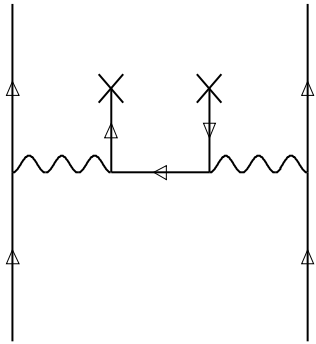
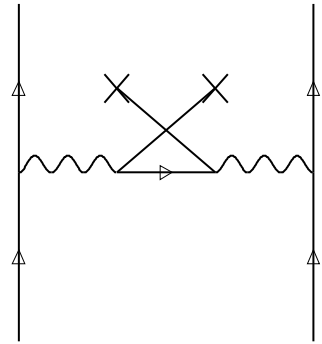
Fig.~1. The Feynman diagrams for the  contributions of  the nonperturbative  
corrections to perturbative quark-quark potential  in  OGE 
approximation with  the lowest dimensional quark and gluon  condensates.  

\vspace{1cm}

\noindent
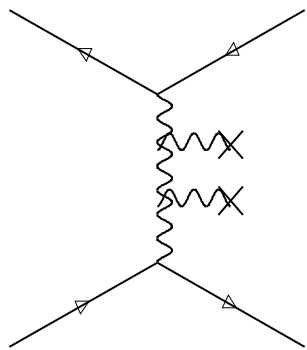
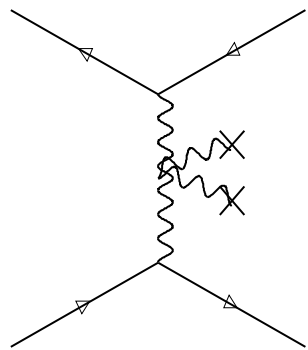
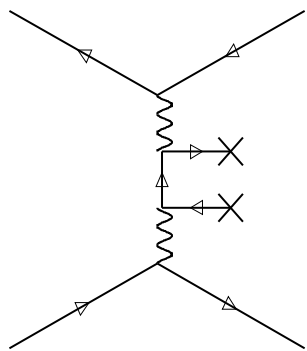
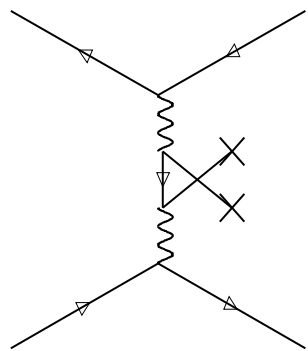
Fig.~2. The Feynman diagrams for the contributions of  the nonperturbative  
corrections to perturbative $q \bar{q}$-pair annihilation potential  in  
OGE approximation with  the lowest 
dimensional quark and gluon  condensates.  

\vspace{1cm}

\noindent
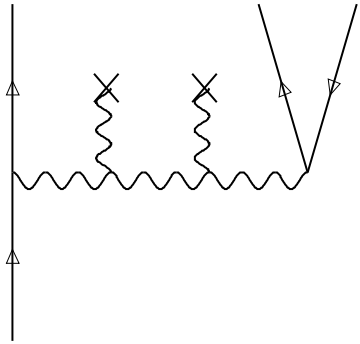
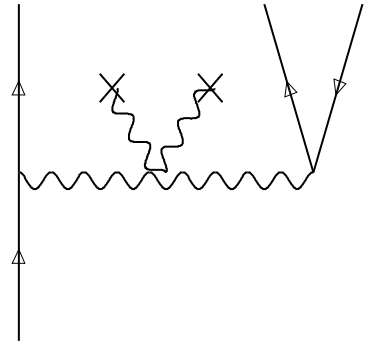
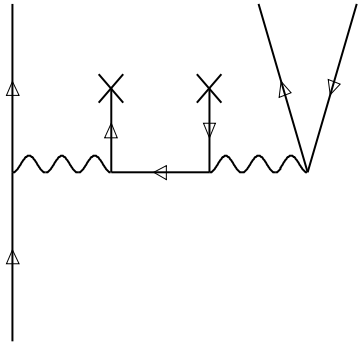
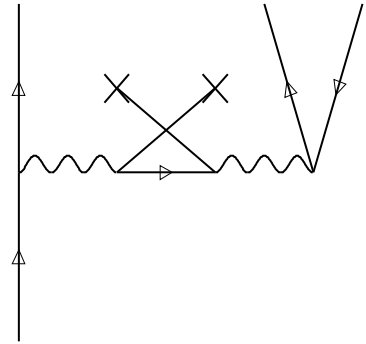
Fig.~3. The Feynman diagrams for the contributions of the nonperturbative  
corrections to perturbative $q \bar{q}$-pair excitation potential  in  
OGE approximation with  the lowest 
dimensional quark and gluon  condensates.

\end{document}